\documentclass{pasj00}
\draft

\begin{document}
\SetRunningHead{J. Nakashima and S. Deguchi}{SiO Maser Survey toward the Inner Galactic Disk}
\Received{2001/07/24}
\Accepted{}

\title{SiO Maser Survey toward the Inner Galactic Disk: \\ $40^{\circ} \leq l \leq 70^{\circ}$ and $|b|\leq 10^{\circ}$}

\author{Jun-ichi \textsc{Nakashima}%
  \thanks{Present address: Department of Astronomy, University of Illinois at Urbana-Champaign, 1002 West Green Street, MC-221, Urbana, IL 61801}
  }
\affil{Department of Astronomical Science, The Graduate University for Advanced Studies,\\ Nobeyama Radio Observatory, Minamimaki, Minamisaku, Nagano 384-1305}
\email{junichi@astro.uiuc.edu}
\and
\author{Shuji \textsc{Deguchi}}
\affil{Nobeyama Radio Observatory, National Astronomical Observatory,\\ Minamimaki, Minamisaku, Nagano 384-1305}\email{deguchi@nro.nao.ac.jp}

%


\maketitle

\begin{abstract}
We present the results of an SiO maser survey for color-selected IRAS sources in the area $40^{\circ}<l<70^{\circ}$ and $|b| < 10^{\circ}$ in the SiO $J=1$--$0$, $v=1$ and $2$ transitions ($\sim$ 43 GHz). We detected 134 out of 272 observed sources in SiO masers; 127 were new detections. A systematic difference in the detection rates between SiO and OH maser searches was found. Especially, in the color ranges with $\log (F_{25}/F_{12})$ smaller than $-0.1$, the detection rate of the SiO masers is significantly higher than that of OH masers. We found a possible kinematic influence of the galactic arm on the distribution of SiO maser sources. It was found that the velocity dispersion of SiO maser sources tends to decrease with the galactocentric distance. Using the present and previous data of SiO maser surveys, we found that the local velocity gradient of the rotational velocity of the Galaxy is consistent with the values obtained from other kinds of disk population stars within a statistical uncertainty. The Oort's constants, \textit{A} and \textit{B}, were computed from the gradient of the rotation curve for the present data, and were consistent with the IAU standard values. In addition, in order to check the reliability of IRAS positions, we observed toward the MSX positions for 5 MSX counterparts, which are located more than \timeform{20''} away (but within \timeform{60''}) from IRAS positions. We detected all of these 5 sources in SiO masers.
\end{abstract}

\section{Introduction}
OH/SiO masers from evolved stars have proven to be powerful tools for investigating stellar motion, even in optically obscured regions in the Galaxy (e.g., \cite{lin92}; \cite{ima02}). Line profiles of OH masers usually show double-peak profiles (e.g., \cite{dav93}). It has been well established that the middle velocity of the OH 1612 MHz double peaks gives the stellar velocity, where they are emitted from the approaching and receding parts of the circumstellar expanding shell (cf., \cite{hab96}). The radial velocity of the SiO maser emission from evolved stars usually falls in the middle of the OH 1612 MHz double peaks, giving the radial velocity of the central star to within a few km s$^{-1}$ accuracy (\cite{jew91}). Up to now, large surveys in OH 1612 MHz have been made, and have covered almost all of the Galactic disk and bulge regions (e.g., \cite{tel91}; \cite{che93}; \cite{blo94}; \cite{sev97}). SiO maser surveys have also been made toward various regions of the Galaxy: bulge regions (Izumiura et al. 1994, 1995a, b), Galactic center regions (\cite{ima02}; \cite{miy01}), outer-disk (\cite{jia96}) and high galactic latitude regions (\cite{ita01}). In past SiO maser surveys, the Nobeyama 45-m telescope has played an important role because of its remarkable sensitivity at 43 GHz. As of today, however, an inner-disk region, $40^{\circ} < l < 70^{\circ}$, has never been surveyed in SiO masers at Nobeyama. 

The inner-disk region provides a chance for us to tackle a couple of interesting themes, if it is surveyed systematically in SiO masers. First, one of the themes is a confirmation of the differences in the natures between OH/IR stars and IRAS/SiO sources. Although both of the OH/SiO masers are known to be indicators of O-rich stars (here, "O-rich" means a C/O ratio less than 1), the overlap of two masers is limited to only one third. This insignificance of the overlap suggests that IRAS/SiO sources should have different physical conditions compared with the OH/IR stars. A precise comparison from various viewpoints is needed to reveal the physical difference of these two objects. Because sensitive OH maser (1612MHz) surveys toward IRAS sources have been made with the Arecibo 300-m telescope in the inner-disk region (e.g., \cite{lew90}), we can make a minute comparison of the nature of OH/IR stars and IRAS/SiO sources using the result of SiO maser survey in the inner-disk region. Second, mass dependency of chemical evolution of evolved stars can be investigated in the inner-disk region. It has been suggested that O-rich AGB stars are predominantly located in the Galactic arms (e.g., \cite{jia93}). This suggestion is attractive in terms of the chemical evolution of AGB stars. According to the dynamics of the Galaxy, AGB stars with a mass larger than 2--3 $M_{\odot}$ are preferentially found in the spiral arms. Therefore, if the concentration of O-rich star into the arms is real, massive AGB stars might prevent a chemical evolution from O-rich to C-rich stars, although the evolution from O-rich to C-rich star is a standard scheme of AGB evolution (here, "C-rich" means C/O ratio larger than 1). Because the spatial concentration of the O-rich AGB stars into the arms is relatively weak, a kinematical study is needed to confirm the affection from the arms. A tangential point of the Sagittarius--Carina arm is located at 5 kpc from the Sun in the inner-disk region; it is a quite appropriate distance to make a precise investigation of SiO maser sources in terms of the sensitivity of the Nobeyama 45-m telescope. (The tangential point means a point where a tangential line through the Sun contacts with the galactic arm.) Third, a local gradient of the rotation curve of the Galaxy is also an interesting theme to be investigated by an SiO maser survey in the inner-disk region. In the outer-disk region ($90^{\circ} < l < 230^{\circ}$), rotational velocities of O-rich AGB stars have been obtained based on the data of past SiO maser surveys (\cite{jia96}; \cite{nak00}). On the other hand, the number of known SiO maser sources is quite limited in the inner-disk region, because of the absence of an SiO maser survey with the Nobeyama 45-m telescope. Therefore, the local inclination of the rotation curve of the Galaxy was never obtained through the motion of IRAS/SiO sources. Although the value has been obtained through a few kinds of disk population stars, a new value of the local inclination of the Galactic rotation curve through new kinds of disk populations would contribute to obtain more precise Galactic constants (e.g., \cite{ker86}).

In order to research the above issues, we systematically surveyed a sample of IRAS sources in the inner-disk region in SiO maser lines. In this paper, we report on the result of this systematic SiO maser survey. The outline of the paper is as follows. In section 2 we describe the observational details, results of observations, statistical characteristics of the data, and the reliability of the IRAS positions. In section 3 we give a statistical comparison with an OH maser survey and a kinematical analysis of the present data, and finally conclude the paper in section 4.

\section{Observations and Results}
\subsection{SiO Maser Survey toward IRAS Sources}
Simultaneous observations in SiO $J=1$--$0$, $v=1$ and $2$ transitions at 43.122 and 42.821 GHz, respectively, were made with the 45-m radio telescope at Nobeyama during the period from 2000 April to 2001 March. The beam size of the telescope was about 40$''$ at 43 GHz. In the present observation sessions, a cooled SIS receiver (S40) with a bandwidth of about 0.4 GHz was used, and the system temperature (including atmospheric noise) was about 200--300 K (SSB). The aperture efficiency of the telescope was about 0.57 at 43 GHz. The conversion factor from the antenna temperature to the flux density was 2.9 Jy K$^{-1}$. Acousto-optical spectrometer arrays of high and low resolutions (AOS-H and AOS-W) were used. The AOS-H spectrometer has 40 MHz bandwidth and 2048 frequency channels with an effective spectral resolution of 0.29 km s$^{-1}$ at 43 GHz. Likewise, the AOS-W spectrometer has a 250 MHz bandwidth and 2048 frequency channels with an effective spectral resolution of 1.7 km s$^{-1}$ at 43 GHz. We used multiple AOS-H spectrometers to cover a wide velocity range at a high frequency resolution. Because line widths of masers are often less than 1 km s$^{-1}$, we needed the high resolution achieved by AOS-H. The radial-velocity coverage was $\pm$ 350 km s$^{-1}$. This velocity range was fully covered by the AOS-H spectrometers (AOS-W was used to confirm the detections). All of the observations were made by the position-switching mode using a 5$'$ off position in the azimuth. The pointing of the telescope was checked every 2 or 3 hours by 5-point mapping of nearby SiO maser sources, V1111 Oph and $\chi$ Cyg. The pointing accuracy was usually found to be better than 10$''$.

We selected a sample of observing sources in the Galactic plane area, $40^{\circ}<l<70^{\circ}$ and $|b|<10^{\circ}$, from the IRAS Point Source Catalog (Version 2, here after PSC). The distribution of the sources in the galactic coordinates is shown in figure 1. The source selection was made in terms of the IRAS 12 $\mu$m flux density, the IRAS colors and measurement quality flags. The criteria of the selection are as follows:
\begin{enumerate}
\item $F_{12} > 3$ Jy,
\item $-0.2 < C_{12} \equiv \log (F_{25}/F_{12}) < 0.2$,
\item $C_{23} \equiv \log (F_{60}/F_{25}) < -0.5$,
\item $Q_{12} = Q_{25} = 3$,
\end{enumerate}
where $F_{12}$, $F_{25}$, and $F_{60}$ are the IRAS 12, 25 and 60 $\mu$m flux densities, $Q_{12}$ and $Q_{25}$ are the measurement quality flags for 12 and 25 $\mu$m, respectively, and "3" means the highest quality. These selection criteria effectively extract dust-enshrouded objects with $T_{\textrm{dust}} \simeq 240$--$450$ K from the IRAS PSC. It was found in past SiO maser surveys that the detection rate of SiO masers is maximized at an IRAS color, $C_{12} \simeq 0$ ($T_{\textrm{dust}} \simeq 300$ K) and that the rate drops quickly in $|C_{12}|>0.2$ (e.g., \cite{jia95}; \cite{nym98}). Therefore, we centered the color of $C_{12} = 0.0$ in the criterion 2 to maximize the detection rate. Finally, 340 sources were selected from IRAS PSC by above criteria, and 272 have been observed. Almost all of the samples above $F_{12} = 8$ Jy were observed (more than 99\%). In the case of faint sources with $F_{12} < 8$ Jy, the completeness of the survey is somewhat low (mentioned later). 

Raw data were processed by flagging out bad scans, making r.m.s.-weighted integrations, and removing the slope in the baseline. The detections of maser lines were judged by the following criteria. For narrow spike-type emissions, the peak antenna temperature must be greater than the 3 $\sigma$ level of the r.m.s. noise. For broad and multiple spiky emissions, the effective signal-to-noise ratio (S/N) over the line width must be larger than 5; the effective S/N is calculated from integrated intensities within maximum line width (about 15 km s$^{-1}$ in typical cases) and the rms noise. In addition, we carefully inspected each spectrum by eye and discarded some spurious emissions and marginal detections that satisfy the above criteria. 

As a result of the present observation, we detected 134 out of 272 observed sources in either SiO $J=1$--$0$, $v=1$ or $2$ transitions; 127 of 134 detected sources were newly detected objects in SiO masers. The observational results are summarized in table 1 for the detected sources and in table 2 for the non-detected sources. The spectra of the detected sources are shown in figure 2. Velocities, V$_{\textrm{lsr}}$, given in table 1 are radial velocities at an intensity peak in the spectra, except for the case of IRAS 19229+1708. The intensity-weighted radial velocity normally coincides with the velocity at the peak within a few km s$^{-1}$ (\cite{jew91}). IRAS 19229+1708 exhibits an extreme broad line-width with triple peaks. In addition, the intensity peak of the $v=1$ line of IRAS 19229+1708 is clearly shifted from its systemic velocity according to the profile of $v=2$ line. Therefore, we adopted an intensity weighted mean of radial velocity as a systemic velocity for IRAS 19229+1708. Finally, the detection rate was 49\%. The detection rates in the ranges of $40^{\circ} < l < 50^{\circ}$, $50^{\circ} < l < 60^{\circ}$, $60^{\circ} < l < 70^{\circ}$ are 61\%, 47\% and 38\%, respectively. This variation of the detection rate with galactic longitude can be seen in figure 1.\\
\\
(Figure 1 here)
\\
(Figure 2 here)
\\
(Table 1 here)
\\
(Table 2 here)
\\

\subsection{Position Confirmation Using the MSX Catalog}
It has been believed that the uncertainties of the source positions in the IRAS PSC are within about \timeform{10''} (\cite{bei88, jia97}). This position uncertainty, if it is reliable, is sufficiently accurate for observations by the Nobeyama 45-m telescope, because of its half-power beam width of \timeform{40"} at 43 GHz. Recently, however, \citet{deg01} found a non-negligible number of near-infrared counterparts of IRAS sources at positions separated from its IRAS positions by more than 20$''$. Because an observing region in \citet{deg01} is close to the galactic center, it seems that the main reasons for the position uncertainty of IRAS PSC are congestion of sources and a large beam of an infrared detector of IRAS. On the contrary, the sources in the present surveyed region were not very crowded compared with the region researched by \citet{deg01}, and the IRAS positions would be relatively reliable than the \citet{deg01}'s case. Here, however, to make sure of the reliability of the source positions of the present sample, we compared the IRAS positions of SiO non-detections with MSX positions (MSX: Midcourse Space Experiment; \cite{pri97}). Because the typical uncertainty of astrometric positions in the MSX catalog is a few arcseconds, the reliability of the IRAS positions could be checked by cross checking with the MSX positions. As a consequence, we found that 5 IRAS sources with SiO non-detection separate from the MSX positions by more than \timeform{20"} (but within \timeform{60"}). In table 3, the separations between the IRAS position and the MSX position, C-band (centered at 12.13 $\mu$m) and E-band (centered at 21.34 $\mu$m) intensities in the MSX catalog, and IRAS 12 and 25 $\mu$m flux densities of these 5 IRAS sources are shown. The MSX flux densities are approximately proportional to the IRAS flux densities, though some scatter is seen. The scatter would be due to the pulsation of AGB stars. We observed these 5 sources once again at their MSX positions, and were detected in SiO masers for all 5 sources. The results of these additional observations are summarized in table 4, and the spectra of the SiO masers are shown in figure 3. In order to maintain consistency with previous SiO maser surveys made at Nobeyama, above 5 detections at the MSX positions were excluded in the statistical analysis discussed in the following sections, because the IRAS positions were used as the observing positions in previous SiO maser surveys. However, in the present research, the number of sources with a large position uncertainty of more than \timeform{20''} was only 5 in 138 nondetections, which would be negligible in the statistics. 
\\
(Figure 3 here)
\\
(Table 3 here)
\\
(Table 4 here)
\\

\subsection{Statistical Characteristics}

In the present work, we extended the survey latitude up to $|b|=10^{\circ}$ ($|b|<3^{\circ}$ in previous works, cf. \cite{izu99}; Deguchi et al. 2000a, b).  Therefore, we checked the variation in the detection rate with the galactic latitude. The detection rates were 47\% in the range of $|b| < 3^{\circ}$ and 52\% in the range of $3^{\circ} < |b| < 10^{\circ}$. No remarkable difference could be seen between these two values. The detection rate of SiO masers, 49\%, in the present survey suggests a decreasing tendency of the detection rate with the galactic longitude. This tendency is consistent with the results in previous SiO maser surveys toward the inner galactic area (\cite{izu99}; Deguchi et al. 2000a, b; \cite{nak02}; \cite{jia96}). However, we have to be cautious of concerning this result, because of the difference in the color selection criteria of each sample. In previous SiO maser surveys towards the inner bulge (\cite{izu99}; Deguchi et al. 2000a, b), the color range, $0.0 < C_{12} < 0.1$, was used as a source selection criterion. In every 10$^{\circ}$-step sub-samples from $l=40^{\circ}$ in the longitude, the detection rates in the color range, $0.0 < C_{12} < 0.1$, are 44\% ($40^{\circ} < l < 50^{\circ}$), 40\% ($50^{\circ} < l < 60^{\circ}$), and 31\% ($60^{\circ} < l < 70^{\circ}$), respectively, in the present survey; the average detection rate was 42\%. In fact, detection rates in previous surveys were 56\% for the bulge bar ($15^{\circ} < l < 25^{\circ}$, $|b| < 3^{\circ}$; \cite{izu99}), 49\% for the inner galactic bulge ($|l| < 3^{\circ}$, $|b| < 3^{\circ}$; \cite{deg00a}a), 62\% for the outer galactic bulge ($-10^{\circ} < l < 25^{\circ}$, $|b| < 3^{\circ}$; \cite{deg00b}b), and 48\% for the inner galactic disk ($25^{\circ} < l < 40^{\circ}$, $|b| < 3^{\circ}$; \cite{nak02}). The present detection rate is less than the values obtained in the inner Galaxy surveys. This clearly confirms the decreasing tendency of the detection rate with the galactic longitude. The detection rate in the outer galactic disk ($90^{\circ} \leqq l < 230^{\circ}$, $|b|<10^{o}$), which is limited in the sample with the color, $0.0 < C_{12} < 0.1$, was 0\% (14 sources were observed). The detection rate obtained in the present work lies between the two detection rates of the outer- and inner-disk surveys, supporting a monotonous decrease of the detection rate with the galactic longitude. All of the SiO surveys mentioned here were made using the same telescope (Nobeyama 45-m telescope), receiver (S40), and observation mode (position-switching). In addition, the average integration times per one object were almost same (about 10 min on source) in all surveys. Therefore, the sensitivities can be considered to be sufficiently similar to compare the detection rates. 

\subsection{IRAS Characteristics of the Sample}

Figure 4 shows a histogram of the IRAS 12 $\mu$m flux densities and the variation of the detection rate with the IRAS 12 $\mu$m flux density in the present sample. The detection rate tends to increase with the 12 $\mu$m flux density. In ranges of $20 < F_{12} < 24$ Jy and $F_{12} < 4$ Jy, the detection rates tend to be slightly higher and lower, respectively, than those of the other ranges. In $20 < F_{12} < 24$ Jy, because the number of samples is extremely few, the rise may be a statistical fluctuation. On the other hand, in $F_{12} < 4$ Jy, because the number of samples in this range is more than 25, it can not be disregarded. In fact, this kind of decline in the detection rate has been seen in previous SiO maser surveys (e.g. \cite{deg00b}b; \cite{izu99}). 

Generally speaking, the intensity of SiO maser lines tends to increase with the IRAS 12 $\mu$m flux density. For instance, in the case of the NRO 45-m telescope, the expected antenna temperature of the SiO maser line from IRAS PSC source with $F_{12}=4$ Jy is approximately 0.2 K. Because of the time variation, the intensity of a maser line varies by a factor of about 10. Therefore, the antenna temperature of the SiO maser line from the IRAS PSC sources with $F_{12}=4$ Jy is expected to lie between about 0.04 and 0.4 K. The r.m.s. level of the present survey data was 0.05 K on average. In the present work, we used the 3 $\sigma$ level as a border of detection (for narrow lines; most of detections falls in this case). The 3 $\sigma$ level corresponds to 0.15 K in the antenna temperature. Hence, it is possible that nearly half of the potential SiO detections in the range of $3 < F_{12} < 4$ Jy are missed in the present survey. If we made exposures for a longer time than in the present case, the detection rate could be increased up to about 40\% in the range of $3 < F_{12} < 4$ Jy according to the detection rates in other intensity ranges.

Figure 5 shows the two-color (left) and $C_{23}$--b (right) diagrams for the present sample. The SiO detection rates were 68\%, 52\%, 50\% and 30\% in the $C_{23}$ ranges of $-0.9 < C_{23} < -0.8$, $-0.8 < C_{23} < -0.7$, $-0.7 < C_{23} < -0.6$, and $-0.6 < C_{23} < -0.5$, respectively. In general, carbon stars in the two-color diagram distribute at the upper part of O-rich stars (\cite{van90}). The present sample involves three previously known carbon stars (IRAS 19029+0808, IRAS 19238+1159, IRAS 19304+2529, \cite{jur90}; \cite{vol92}), as indicated in the left panel by the large crosses. The three carbon stars distribute at the uppermost part in the left panel. This fact supports that the rate of carbon stars in the sample tends to increase with an increase of the color $C_{23}$. 

In the case of the previous SiO maser surveys towards inner-disk/bulge regions, $C_{23}$--$b$ diagram showed an arch-like structure at the bottom (the peak of the arch is at $b=0^{\circ}$), indicating a strong reddening in the Galactic plane (e.g. Figure 4 of \cite{deg00b}b). On the contrary, such a structure cannot be seen in the $b$--$C_{23}$ diagram in the present sample. This would suggest that the contamination by the interstellar dust in the $C_{23}$ color is not strong in the present sample, although the present criteria of source selection is slightly different from those in previous surveys. \\
\\
(Figure 4 here)
\\
(Figure 5 here)
\\

\subsection{Comparison with OH Data}
Figure 6 shows a histogram of the color, $C_{12}$, for the observed sources. The number of sources surveyed in both SiO and OH masers was 240 (in 272 sources), and the number of sources, which were detected in both of SiO and OH masers, was 75. Mainly, OH maser surveys were made with the Arecibo 300-m telescope (about 95\% of OH observations in table 5 were made at Arecibo). Owing to the large aperture of the Arecibo 300-m telescope, the OH maser survey was made deeper than the present SiO maser survey. The shaded portions show the numbers of SiO and OH detections. The solid and dotted lines (line graph) show the detection rates of the present SiO maser survey and the OH maser survey.

The main characteristic in figure 6 is a difference of the two line graphs. The detection rate of the SiO maser is roughly constant through all of the color ranges. On the other hand, the detection rate of the OH maser tends to increase with the color. The present survey is somewhat incomplete for sources with $F_{12} < 4$ Jy. However, the above tendency does not change even if the sources with $F_{12} < 4$ Jy are excluded. This difference of the detection rates between the OH and SiO masers implies that evolved AGB stars with a thick circumstellar envelope tend to exhibit OH masers more often than SiO masers, though AGB stars with a thin circumstellar envelope tend to exhibit SiO masers more than OH masers. However, large ($\gtrsim$ 80\%) OH detection rate may be due to poor statistics at $C_{12}>0.1$. 

We should thus note any difference in the variability of OH/SiO masers. In this kind of survey, the detection rate of SiO masers always gives a lower limit for the true probability for the appearance of SiO masers because of their strong variability. Whereas OH masers are relatively stable compared to SiO, they are also variable. If we make additional surveys for non-detections in SiO, the detection rate would increase in some measure. This fact supports the higher detection rate of SiO than that of OH in the bluer color range in figure 6. The IRAS color, $C_{12}$, has been believed to be an evolutional indicator of AGB stars. The value of $C_{12}$ tends to increase with the evolution of AGB stars (e.g., \cite{van88}). In this sense, OH masers tend to be appendant in more evolved objects than SiO masers. \\
\\
(Figure 6 here)
\\
(Table 5 here)
\\

\section{Discussion}
\subsection{Luminosity Distance}
The distance is calculated from the bolometric flux of the sources, which is computed from the IRAS 12 $\mu$m flux density, $F_{12}$, and a bolometric correction for O-rich stars (\cite{van89}). In the present work, we assumed that the luminosity of the AGB star is 8000 $L_{\solar}$; this corresponds to the luminosity of a star with mass of $\sim$ 2 $M_{\odot}$ near to the tip of the AGB in the models of \citet{vas93}. Details of the distance calculation can be found in our previous papers (e.g., \cite{nak00}). The obtained distances are given in table 5. The luminosity distance involves an uncertainty of about 20--30\%, which is partly due to the uncertainty of the luminosity assumption of 8000 $L_{\solar}$, and partly due to the light variation of AGB stars. In fact, a non-negligible number of color-selected AGB star candidates are mira-type variables (e.g., \cite{nak00}). In the case of short-period miras with $P \lesssim 400$ d, if we know the pulsation period of the star, their distances can be obtained with higher accuracy using the period-luminosity relation (e.g., \cite{nak00}).

\subsection{Kinematical Characteristics of the Sample}

Figure 7 shows $V_{\textrm{lsr}}$ versus the luminosity distance ($D_{L}$) for 134 SiO detected sources. Sources above and below $l=50^{\circ}$ are classified by open and filled circles, respectively. The solid and broken curves are radial velocities expected from the circular motion and the flat rotation curve on the directions $l=40^{\circ}$ and $70^{\circ}$ (see the caption of figure 7 for the details of the flat rotation curve). These two curves gradually separate from each other with the distance. Therefore, if the motion of SiO maser sources follows the galactic rotation (circular motion and flat rotation curve), the open circles ($l<50^{\circ}$) in figure 7 should distribute in a relatively higher velocity range than the open circles ($l>50^{\circ}$). Actually, we can see the difference in the two distribution of the open and filled circles in Figure 7.\\
\\
(Figure 7 here)
\\

Figure 8 shows the $l$--$V_{\textrm{lsr}}$ diagram for SiO detected sources overlaid on the CO $J=1$--$0$ line map (taken from \cite{dam87}). In figure 8, the distribution of the present data is almost the same as that of CO line; this suggests that the present survey reaches to the tangential point. The $V_{\textrm{lsr}}$ distribution of SiO maser sources is relatively wider than the extent of the CO map. This difference in the $V_{\textrm{lsr}}$ distribution would originate mainly in the difference of the velocity dispersions between CO clouds and SiO maser sources. 

A moderate concentration of sources can be seen in the area, $l \simeq 40^{\circ}$--$53^{\circ}$ and $V_{\textrm{lsr}} \simeq 30$ km s$^{-1}$--$60$ km s$^{-1}$. \citet{tay93} pointed out in a study of the distribution of free electrons that the tangential point of the Sagittarius--Carina arm exists at a distance 4--8 kpc from the Sun on the direction of $l=45^{\circ}$. Therefore, it is possible that this weak concentration of sources originates from the Sagittarius--Carina arm (though it is not so clear). A similar weak concentration of sources, possibly due to the Scutum--Crux arm, was also reported in \citet{nak02}. 

We can see a source-deficient region (roughly surrounded by the dotted ellipse in figure 8). In order to confirm this deficiency, we applied the Kolmogorov--Smirnov (K--S) test (\cite{cha67}) for three subsets of the SiO detections divided by the galactic longitude: group A ($40^{\circ}<l<45^{\circ}$), group B ($45^{\circ}<l<50^{\circ}$), and group C ($50^{\circ}<l<55^{\circ}$). The source-deficient region mentioned above is mainly included in group B. The cumulative probability distributions for the three groups are shown in figure 9. The maximum difference in the cumulative probability (D value) between groups A and B, groups B and C, groups C and A are 0.21, 0.24, and 0.13, respectively. The values of $D_{\alpha}$, which provide a boundary to judge at the 99\% confidence level whether the difference of two distributions is reliable, are 0.15, 0.13, and 0.15 for pairs of groups A and B, groups B and C, groups C and A, respectively. According to the D and $D_{\alpha}$ values, the difference in the distribution along the perpendicular axis in figure 9 is significant with more than a 99\% confidence level between groups A and B and groups B and C, but not between groups A and C. In this statistical treatment with K--S test, we assumed that the sampling condition was unity around the source--deficient region. The maximum different points in the cumulative probability diagram between groups A and B and between groups B and C lie just at the source-deficient region. Based on these results, we conclude that the source-deficient region surely exists on the $l$--$v$ diagram. \\
\\
(Figure 8 here)
\\
(Figure 9 here)
\\

\subsection{Variation in the Velocity Dispersion}
We divided the detected sources into three groups based on the galactocentric distance: I ($R < 6.5$ kpc, 25 sources), II ($6.5 \leqq R < 7.5$ kpc, 64 sources), and III ($R \geqq 7.5$ kpc, 44 sources). Then, the velocity dispersion (root mean square of the residues, which is the result of subtracting the circular and solar motions from $V_{\textrm{lsr}}$) was calculated for three groups. Here, we assumed that the circular orbit and flat rotation curve as the galactic rotation (see the caption of figure 7 in detail), and that the galactocentric distance of the Sun is 8.5 kpc. The calculated velocity dispersions are 35.2 ($+$8.2, $-$5.2) km s$^{-1}$, 32.2 ($+$4.3, $-$3.2) km s$^{-1}$, and 30.6 ($+$5.1, $-$3.6) km s$^{-1}$ for groups I, II, and III, respectively. The values in parenthesis are the 80\% confidence intervals. It seems that the velocity dispersion tends to decrease with the galactocentric distance. In order to confirm this tendency from a statistical viewpoint, an F-test was made. The confidence level of the difference in the dispersion between the nearest group I ($R<6.5$ kpc) and the furthest group III ($R \geqq 7.5$ kpc) is 79\%. According to the results from the F-test, we cannot rule out a possibility for the increasing tendency of the velocity dispersion with the galactic radius, though the statistical confidence level is somewhat low. It should be noted that the velocity dispersions obtained here are a somewhat weighted mean of the radial and tangential dispersions, with a higher weight to the tangential one. This effect is most significant in group III. In addition, the velocity dispersion obtained here is similar to that for M-type stars in the solar neighborhood (e.g., \cite{deh98}; \cite{mih81}).

\subsection{Influence of Arms on the SiO Maser Source Distribution}
In the previous subsection 3.2, although we found that the weak concentration of SiO sources in the $l$--$V_{\textrm{lsr}}$ diagram (figure 8) is possibly due to the spiral arm, it was not so clear. In order to confirm the spatial distribution of the present sample, the positions of the present sample projected onto the galactic plane are shown in figure 10. A spiral model, which is indicated as the background in figure 10, is referred from \citet{tay93}. In figure 10, the IRAS sources in the present sample ($F_{12} > 3$) for which observations were not made are also plotted with small dots. In addition, faint sources with $F_{12}=1$--3 Jy, which have the same colors as those in the present sample, are plotted with small crosses. Because the present survey is incomplete for faint sources with $F_{12} < 3$ Jy, the distribution of the observed sources is inhomogeneous in a distant region over 6 kpc from the Sun. On the contrary, the distribution of the observed sources in the nearby region within 6 kpc is relatively homogeneous because of the completeness of the present survey for bright sources with $F_{12} > 3$ Jy. 

Although the distribution of all samples is almost homogenous, except for the distant region, the individual distributions of open/filled circles are systematically different from each other in figure 10. Specifically speaking, the number density of the filled circles tends to increase with a decrease of the galactic longitude. It is difficult to say based only on the present data that this tendency originates from the influence of the galactic arm, but we should note that the tangential point of the Sagittarius--Carina arm exist in the direction of $l \sim 40^{\circ}$, where the number density of filled circles (SiO detections) is maximized. \citet{jia93} suggested that the spatial distribution of O-rich AGB stars shows a similar spiral structure as that presented by \citet{geo76}. Fitting with the spiral structure of \citet{geo76}, \citet{jia93} derived that the average luminosity of the objects in the sample is 8300 $L_{\odot}$; this value is appropriate as a luminosity of an O-rich AGB star (e.g., \cite{vas93}). This tendency is attractive for studying the chemical evolution of AGB stars, especially with a massive main-sequence mass larger than 3 $M_{\odot}$. According to the dynamics of the Galaxy, AGB stars with masses larger than $\sim 3$ $M_{\odot}$ are preferentially found in the spiral arms (\cite{jia93}). From the concentration of SiO masers, if it is real, most SiO maser sources in the concentration might be O-rich massive AGB stars with the main-sequence mass being larger than $M_{\textrm{ms}} \sim 3$ $M_{\odot}$. This agrees with the current stellar evolution models based on Hot Bottom Burning (nuclear processing of carbon into oxygen and nitrogen at the base of the convective mantle for the most massive AGB stars; \cite{ibe83}; \cite{woo83}). \citet{ven99} show that stars with $M_{\textrm{ms}} > 3.8$ $M_{\odot}$ always keep the O-rich chemistry through their evolution, although AGB stars with M$_{ms} < 3.8$ M$_{\odot}$ evolve from an O-rich star to a C-rich star. In fact, the higher mass limit for carbon stars in the LMC is currently believed to be $\sim 4$ $M_{\odot}$ (\cite{gro93}; \cite{mar99}). In order to clarify whether the concentration of SiO masers into the arms is real or not, we need more accurate information about the distance. Fortunately, in the near future, distances to SiO masers will be directly determined by measuring the annual parallax using Japanese differential VLBI network, VERA (\cite{hon00}).\\
\\
(Figure 10 here)
\\

\subsection{Rotation Curve}
Figure 11 is a plot of the rotational velocity of SiO sources against the galactocentoric distance. The rotational velocity was calculated from the radial velocities by subtracting the rotational velocity (220 km s$^{-1}$) of the local standard of rest, while assuming that the radial velocity is a projection of the circular rotational velocity along the line of sight. The solid line shows the running mean of $V_{\textrm{lsr}}$ in every 1 kpc width with 0.5 kpc steps from 5.5 kpc to 11 kpc. The dotted line shows the best fit line to the data. In the calculation, we added the data in \citet{jia96} in the outer-disk; the selection criteria of the sample and the method to obtain distances in \citet{jia96} are the same as those used in the present work. The least square method gives the best fit,
\begin{equation}
   V_{\textrm{rot}} = 202.2 (\pm 14.24 )\, [{\rm km\,\, s^{-1}}] - 4.8 (\pm 1.7)\, [{\rm km\,\, s^{-1} kpc^{-1}}] \times (R - R_{0}) [{\rm \,kpc}],
\end{equation}
where $R_{0}$ is assumed to be 8.5 kpc. The obtained inclination is consistent with the values obtained for the other kind of disk population stars [$-2.1 \pm 2.4$ km s$^{-1}$ kpc$^{-1}$ for OB stars, $-5.7 \pm 3.2$ km s$^{-1}$ kpc$^{-1}$ for Cepheid (e.g. \cite{fri96})] within a statistical uncertainty. Oort's constants estimated from this inclination are $A=15.4 \pm 0.9$ km s$^{-1}$ kpc$^{-1}$ and $B=-10.5 \pm 0.9$ km s$^{-1}$ kpc$^{-1}$. These values are close to the IAU standard values ($A=15$ km s$^{-1}$ kpc$^{-1}$, $B=-10$ km s$^{-1}$ kpc$^{-1}$; \cite{ker86}). The value of 202.2 km s$^{-1}$ for the rotation velocity at the solar circle is quite similar to those derived from motion of the H{\sc i} gas component by \citet{mer92} and \citet{hon97}. According to the present results, the bulk motion of the SiO maser sources in the solar neighborhood would be the same as that of the other disk population stars and the gas component. 

In recent works on the kinematics of SiO maser sources in the outer-disk, the obtained inclinations of the rotation curve, which were limited to the range of $R=8.5$--11.5 kpc, were $-21.7 \pm 5.7$ km s$^{-1}$ kpc$^{-1}$ (\cite{jia96}), and $-15.4 \pm 7.7$ km s$^{-1}$ kpc$^{-1}$ (\cite{nak00}). These values are somewhat steeper than the present results, indicating that the rotational velocity of the Galaxy tends to decrease in the outer region further than the Sun; this result is consistent with observations of the other kind of disk population stars \citep{fri96}.\\
\\
(Figure 11 here)
\\

\section{Conclusion}
We observed 272 color-selected IRAS sources in the galactic disk area, $40^{\circ} < l < 70^{\circ}$ and $|b| < 10^{\circ}$, in the SiO $J=1$--$0$, $v=1$ and $2$ transitions, and detected 134 in SiO masers; 127 were new detections in SiO masers. The main results of this research are as follows:
\begin{enumerate}

\item A systematic difference in the detection rates between SiO and OH maser searches was found. In the ranges below $C_{12}=-0.1$, the SiO detection rate is higher than the OH detection rate.

\item Possible concentrations of SiO sources in the $l-V_{\textrm{lsr}}$ diagram and in the face-on view diagram might be attributed to the spiral arm (Sagittarius--Carina arm). The observed sources with SiO masers moderately concentrate into the tangential point of the Sagittarius--Carina arm in the spatial distribution.

\item The velocity dispersion of SiO maser sources tends to decrease with an increase of tge galactocentric distance.

\item The data in the present and previous SiO maser surveys give a local inclination of the rotation curve which is consistent with values obtained previously in observations of the other kind of disk population stars and a H{\sc i} gas component.
\end{enumerate}

The authors thank the staff of the Nobeyama radio observatory for their help during observations, and also thank Y. Ita and T. Soma for their help in the data reductions. The authors also thank an anonymous referee, who gave many useful comments and suggestions. This research has made use of the SIMBAD database operated by CDS and MSX database in IPAC. This paper is a part of the JN's thesis presented for the award of Ph. D. to the Graduate University for Advanced Studies.

\appendix
\section{Individual Sources}
\subsection{IRAS 18413+1354 (V837 Her)}
An SiO maser from this source is extremely strong (12 K in the $J=1$--$0$, $v=1$ transition). This source is included in the catalog of nearby dusty AGB stars within 1 kpc \citep{jur89}. This source is also very bright at 12 $\mu$m (225 Jy) with IRAS LRS (IRAS Low Resolution Spectrum) class of 29. OH and H$_2$O masers have been detected (\cite{slo85}; \cite{com90}); $^{12}$CO (1--0) was detected \citep{marg90}; CS (1--0) and NH$_3$ (1, 1) searches were negative \citep{ang96}.

\subsection{IRAS 18530+0817 (EIC 719)}
An IRAS LRS of this source exhibits an unusual spectrum (LRS class 04; 04 means strange band shape). It has strong silicate emission peaking near 9 $\mu$m (\cite{oln86}), which is either self-absorption at 10 $\mu$m or the silicate feature affected by the other molecular bands. SiO masers were detected in the present work. An OH search was negative \citep{lew90}. This object has the strongest stellar photospheric H$_2$O absorption, yet observed in the near-infrared \citep{wal97}.

\subsection{IRAS 18581+1405}
The observed radial velocity, $V_{\textrm{lsr}}=124.4$ km s$^{-1}$, of this source is somewhat high for a disk population star (highest in figures 7 and 8). The main-line of OH maser (1665 MHz) was detected at $V_{\textrm{lsr}}= 114.4$ km s$^{-1}$ \citep{lew97}, which is consistent with the SiO radial velocity. A H$_2$O maser search was negative \citep{eng96}.

\subsection{IRAS 18585+0900}
The IRAS LRS class of this source is 41 (indicating a carbon star). However, this is due to absorption at the 10 $\mu$m silicate feature. In fact, SiO maser emission (only the $J=1$--$0$, $v=2$ transition) was detected in the present work. Double peaks of the OH maser (1612 MHz) gave an expansion velocity of 61 km s$^{-1}$ \citep{ede88}, whereas the expansion velocity in the CO observation was 20 km s$^{-1}$ \citep{lou93}. The H$_2$O maser was negative \citep{eng96}. 

\subsection{IRAS 19192+0922 (AFGL 2374)}
Based of the radio observations of $^{12}$CO ($J=1$--$0$ and $J=2$--$1$ transitions; \cite{hes90}), the mass-loss rate of this source was estimated to be $9.0 \times 10^{-6}$ $M_{\solar}$ yr$^{-1}$. The distance to this source, 1.13 kpc, was obtained by an OH phase lag measurement \citep{vanl90}. The distance (1.6 kpc) used in the present work is consistent with this value. An IRAS LRS class is 31. \citet{mei99} selected this source as a proto-planetary nebula candidate. Because we detected SiO masers in this star, this source might still be in the AGB phase.

\subsection{IRAS 19229+1708}
The profile of SiO masers of this source is somewhat unusual, exhibiting a relatively wide line width and multiple peaks in the $J=1$--$0$, $v=1$ line, which is stronger than that of $v=2$. A H$_2$O maser has been detected \citep{eng96}. 

\subsection{IRAS 19371+2855 (NSV 12260)}
This star was classified as an S-type star, based on the presence of the (0,0) sub-bands of the optical region of LaO \citep{ste90}, but reclassified by \citet{llo99} as a late M-type star. SiO maser detection in the present work and a strong silicate emission (\cite{llo99}) are consistent with the non-S-type. The thermal lines of CO (1--0 and 2--1) were also detected at $V_{\textrm{lsr}} \simeq 20$ km s$^{-1}$ \citep{gro98}, which is consistent with $V_{\textrm{lsr}}=24.3$ km s$^{-1}$ in the present work. OH and H$_2$O have been detected (\cite{che93}; \cite{eng96}).

\subsection{IRAS 19558+3333}
This is a suspected symbiotic star. Radio continuum emission from this source implies a hot, ionizing companion \citep{sea94}. A search for the 3.1 $\mu$m absorption feature, a  characteristics of carbon stars, was negative \citep{gro94}. SiO (present work) and OH \citep{les92} maser searches were negative.

\subsection{IRAS 20056+1834 (QY Sge)}
The optical counter part of this source is a G0 supergiant with a visual magnitude of 12.5,  indicating a post-AGB star. The spectrum has very strong, broad Na D emission lines \citep{men88}. An SiO maser search was negative in the present work.


\newpage
\renewcommand{\thefigure}{1}
\begin{figure}
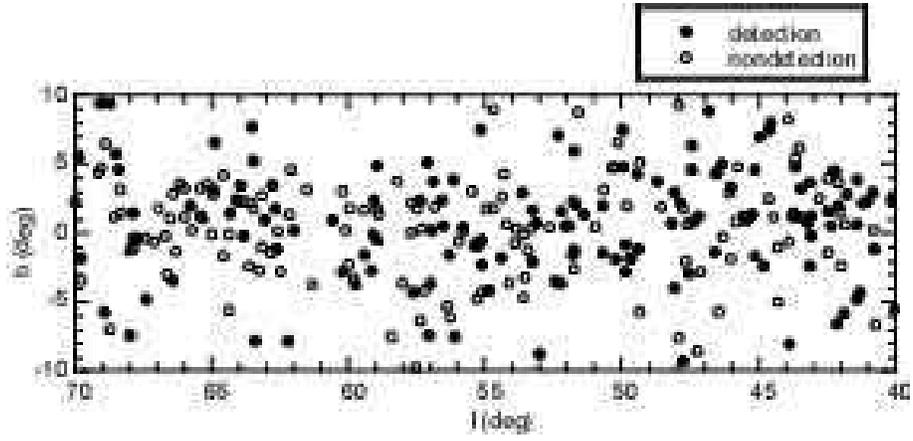

  \begin{center}
    \FigureFile(120mm,10mm){fig1.eps}
  \end{center}
  \caption{Distribution of observed sources in the galactic coordinates. The filled and open circles indicate the SiO detection and non-detection, respectively.}\label{fig:sample}
\end{figure}

\newpage
\renewcommand{\thefigure}{2a}
\begin{figure}
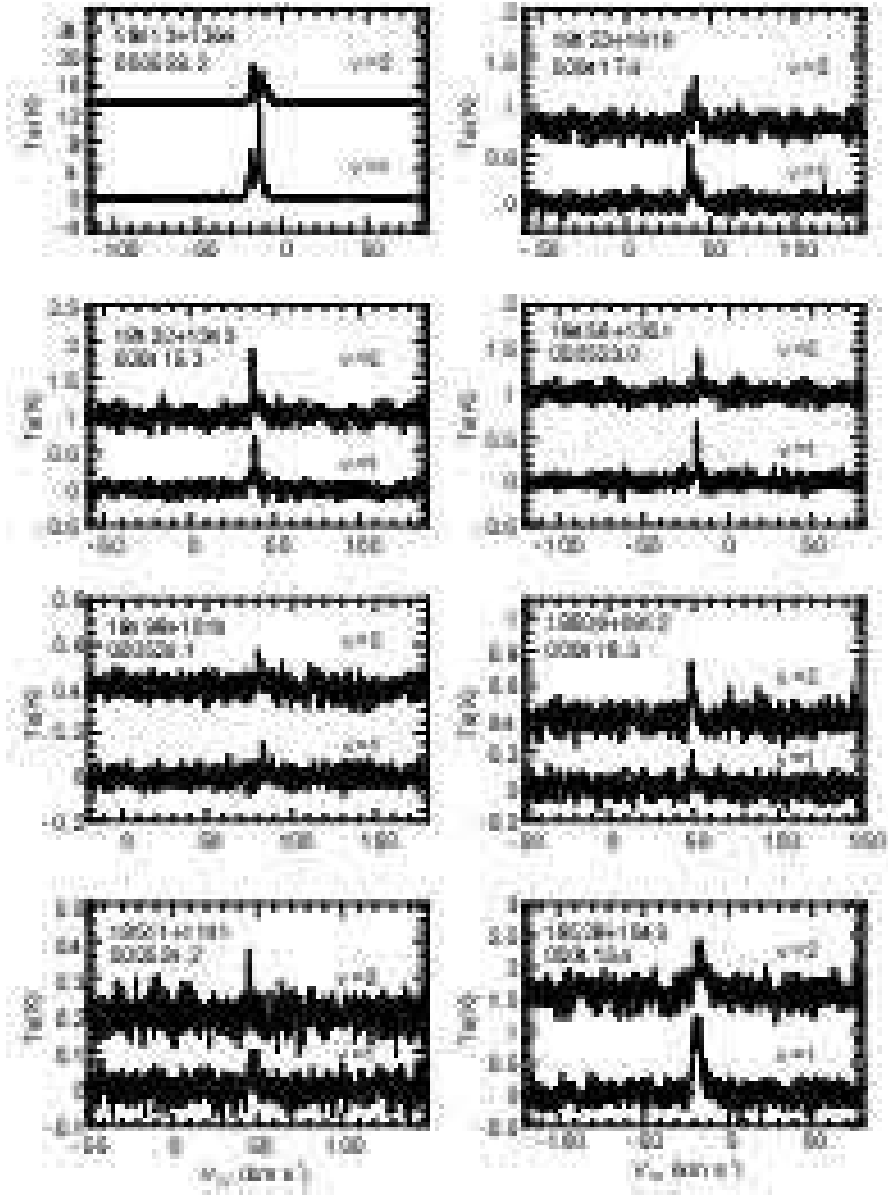

  \begin{center}
    \FigureFile(120mm,10mm){fig2a.eps}
  \end{center}
  \caption{Spectra of the SiO $J=1$--$0$, $v=1$ and $2$ lines for 134 detected sources.}\label{fig:sample}
\end{figure}

\newpage
\renewcommand{\thefigure}{2b}
\begin{figure}
  \begin{center}
    \FigureFile(120mm,10mm){fig2b.eps}
  \end{center}
  \caption{Continued.}
  \label{fig:sample}
\end{figure}

\newpage
\renewcommand{\thefigure}{2c}
\begin{figure}
  \begin{center}
    \FigureFile(120mm,10mm){fig2c.eps}
  \end{center}
  \caption{Continued.}
  \label{fig:sample}
\end{figure}

\newpage
\renewcommand{\thefigure}{2d}
\begin{figure}
  \begin{center}
    \FigureFile(120mm,10mm){fig2d.eps}
  \end{center}
  \caption{Continued.}
  \label{fig:sample}
\end{figure}

\newpage
\renewcommand{\thefigure}{2e}
\begin{figure}
  \begin{center}
    \FigureFile(120mm,10mm){fig2e.eps}
  \end{center}
  \caption{Continued.}
  \label{fig:sample}
\end{figure}

\newpage
\renewcommand{\thefigure}{2f}
\begin{figure}
  \begin{center}
    \FigureFile(120mm,10mm){fig2f.eps}
  \end{center}
  \caption{Continued.}
  \label{fig:sample}
\end{figure}

\newpage
\renewcommand{\thefigure}{2g}
\begin{figure}
  \begin{center}
    \FigureFile(120mm,10mm){fig2g.eps}
  \end{center}
  \caption{Continued.}
  \label{fig:sample}
\end{figure}

\newpage
\renewcommand{\thefigure}{2h}
\begin{figure}
  \begin{center}
    \FigureFile(120mm,10mm){fig2h.eps}
  \end{center}
  \caption{Continued.}
  \label{fig:sample}
\end{figure}

\clearpage
\newpage
\renewcommand{\thefigure}{2i}
\begin{figure}
  \begin{center}
    \FigureFile(120mm,10mm){fig2i.eps}
  \end{center}
  \caption{Continued.}
  \label{fig:sample}
\end{figure}

\newpage
\renewcommand{\thefigure}{2j}
\begin{figure}
  \begin{center}
    \FigureFile(120mm,10mm){figjb.eps}
  \end{center}
  \caption{Continued.}
  \label{fig:sample}
\end{figure}

\newpage
\renewcommand{\thefigure}{2k}
\begin{figure}
  \begin{center}
    \FigureFile(120mm,10mm){fig2k.eps}
  \end{center}
  \caption{Continued.}
  \label{fig:sample}
\end{figure}

\newpage
\renewcommand{\thefigure}{2l}
\begin{figure}
  \begin{center}
    \FigureFile(120mm,10mm){fig2l.eps}
  \end{center}
  \caption{Continued.}
  \label{fig:sample}
\end{figure}

\newpage
\renewcommand{\thefigure}{2m}
\begin{figure}
  \begin{center}
    \FigureFile(120mm,10mm){fig2m.eps}
  \end{center}
  \caption{Continued.}
  \label{fig:sample}
\end{figure}

\newpage
\renewcommand{\thefigure}{2n}
\begin{figure}
  \begin{center}
    \FigureFile(120mm,10mm){fig2n.eps}
  \end{center}
  \caption{Continued.}
  \label{fig:sample}
\end{figure}

\newpage
\renewcommand{\thefigure}{2o}
\begin{figure}
  \begin{center}
    \FigureFile(120mm,10mm){fig2o.eps}
  \end{center}
  \caption{Continued.}
  \label{fig:sample}
\end{figure}

\newpage
\renewcommand{\thefigure}{2p}
\begin{figure}
  \begin{center}
    \FigureFile(120mm,10mm){fig2p.eps}
  \end{center}
  \caption{Continued.}
  \label{fig:sample}
\end{figure}

\newpage
\renewcommand{\thefigure}{2q}
\begin{figure}
  \begin{center}
    \FigureFile(120mm,10mm){fig2q.eps}
  \end{center}
  \caption{Continued.}
  \label{fig:sample}
\end{figure}

\newpage
\renewcommand{\thefigure}{3}
\begin{figure}
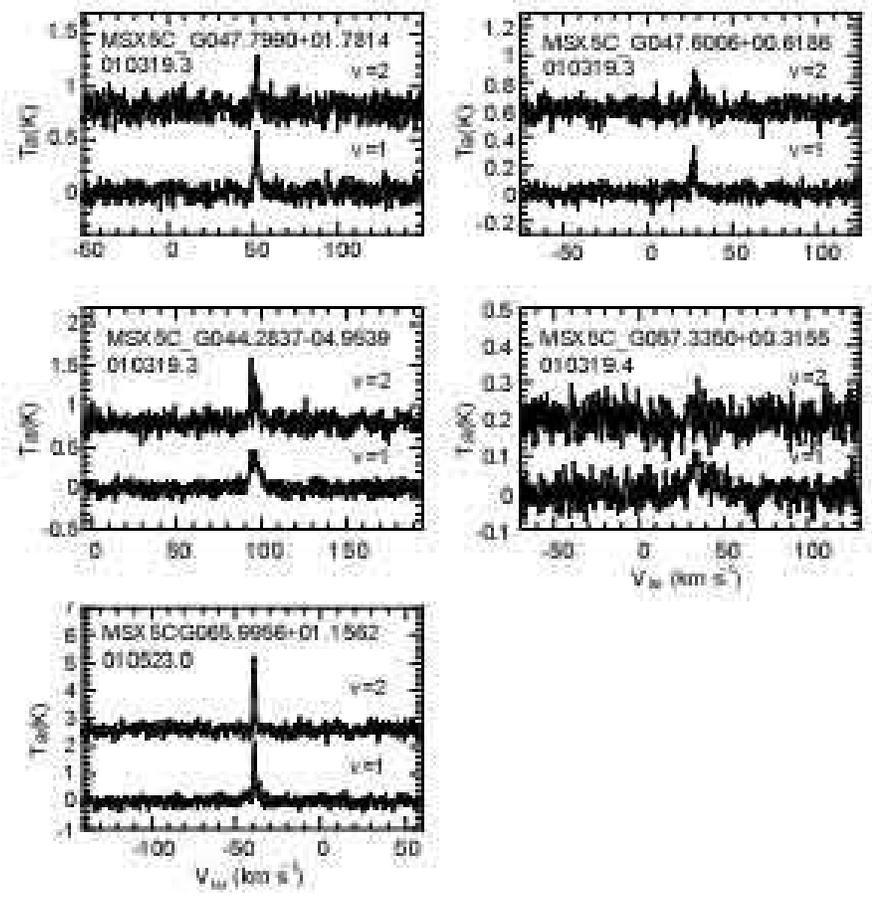

  \begin{center}
    \FigureFile(120mm,10mm){fig3.eps}
  \end{center}
  \caption{Spectra of the SiO $J=1$--$0$, $v=1$ and $2$ lines for MSX sources as counterparts of IRAS PSC with large position uncertainties.}\label{fig:sample}
\end{figure}

\newpage
\renewcommand{\thefigure}{4}
\begin{figure}
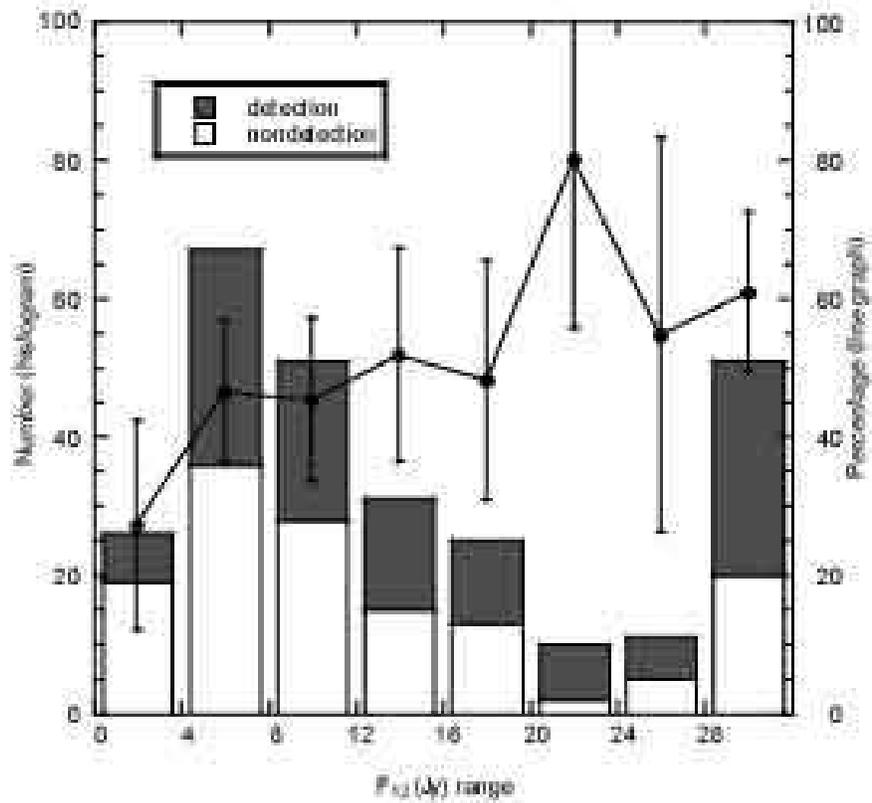

  \begin{center}
    \FigureFile(120mm,10mm){fig4.eps}
  \end{center}
  \caption{Histogram of the 12 $\mu$m flux density (bar graph) and detection rate as a function of the IRAS 12 $\mu$m flux density (line graph). Error bars in the line graph mean 90\% confidence intervals.}\label{fig:sample}
\end{figure}

\newpage
\renewcommand{\thefigure}{5}
\begin{figure}
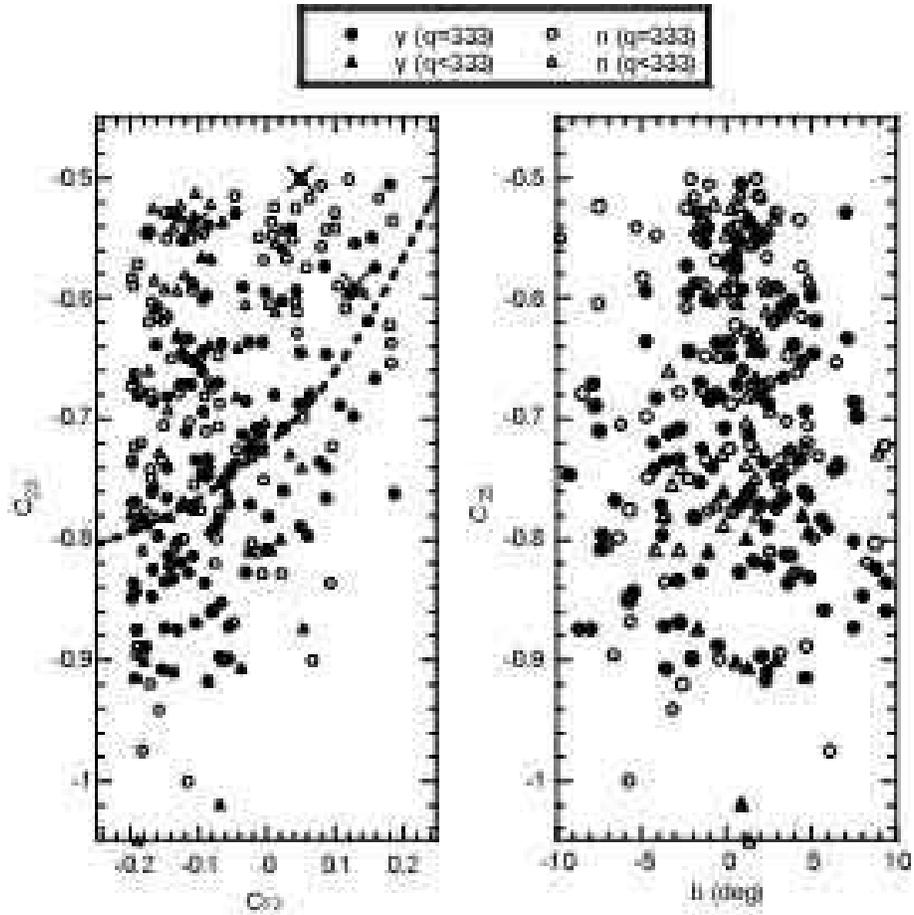

  \begin{center}
    \FigureFile(120mm,10mm){fig5.eps}
  \end{center}
  \caption{Two color ($C_{23}$--$C_{12}$; left panel) and color-latitude ($C_{23}$--$b$; right panel) diagram for the observed sources. The dotted line indicates an evolutionary track of the O-rich AGB star (\cite{van88}). }\label{fig:sample}
\end{figure}

\newpage
\renewcommand{\thefigure}{6}
\begin{figure}
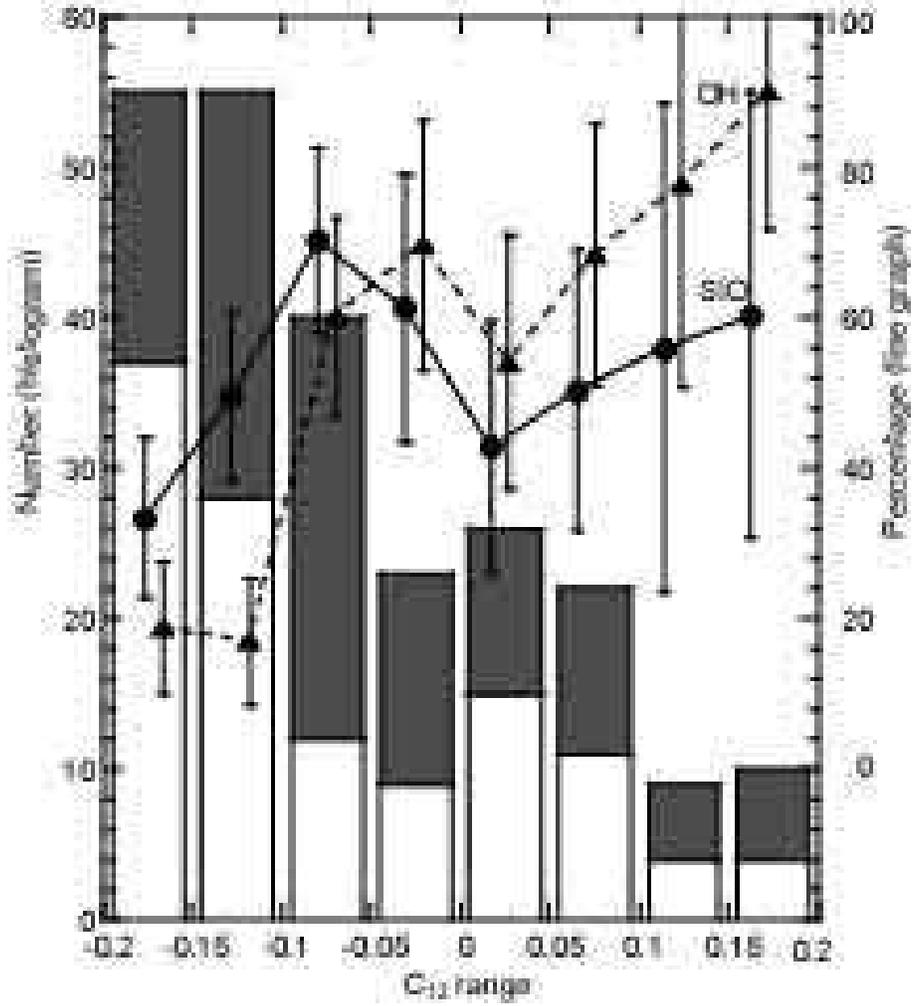

  \begin{center}
    \FigureFile(120mm,10mm){fig6.eps}
  \end{center}
  \caption{Histogram of the color index, $C_{12}$ (bar graph), and detection rate as a function of the $C_{12}$ for SiO maser search (solid line) and OH maser searches (dotted line). Shaded portions in columns indicate the number of detections of SiO and OH masers. Error bars in line graphs mean 90\% confidence intervals. }\label{fig:sample}
\end{figure}

\newpage
\renewcommand{\thefigure}{7}
\begin{figure}
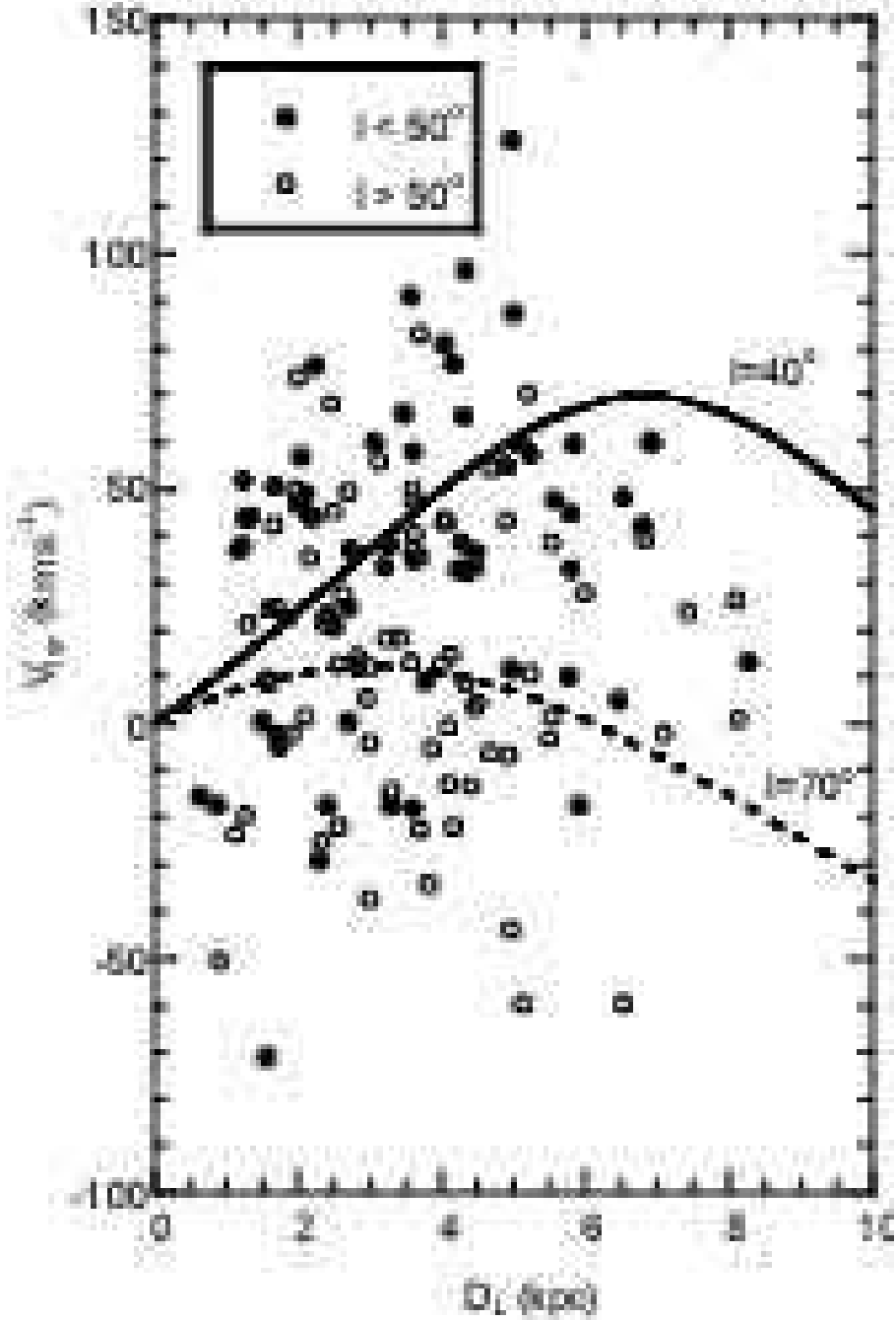

  \begin{center}
    \FigureFile(120mm,10mm){fig7.eps}
  \end{center}
  \caption{Observed radial velocity versus the luminosity distance. The solid and dotted curves are the radial velocities expected at $l=40^{\circ}$ and $70^{\circ}$, respectively, from the Galactic rotation curve $V_{\textrm{rot}} = 220 {\rm km\, s^{-1}} R(8.5 + R_{\textrm{a}})/[8.5(R + R_{\textrm{a}})] \{ 1 + 1.1/[1 + 3(R - R_{\textrm{a}})^2]\}$, where \textit{R} is the distance from the Galactic center in kpc and $R_{\textrm{a}}$ is an adjustable parameter (assumed to be 0.3 kpc) that creates a peak of the rotation curve near $R_{\textrm{a}}$.}\label{fig:sample}
\end{figure}

\newpage
\renewcommand{\thefigure}{8}
\begin{figure}
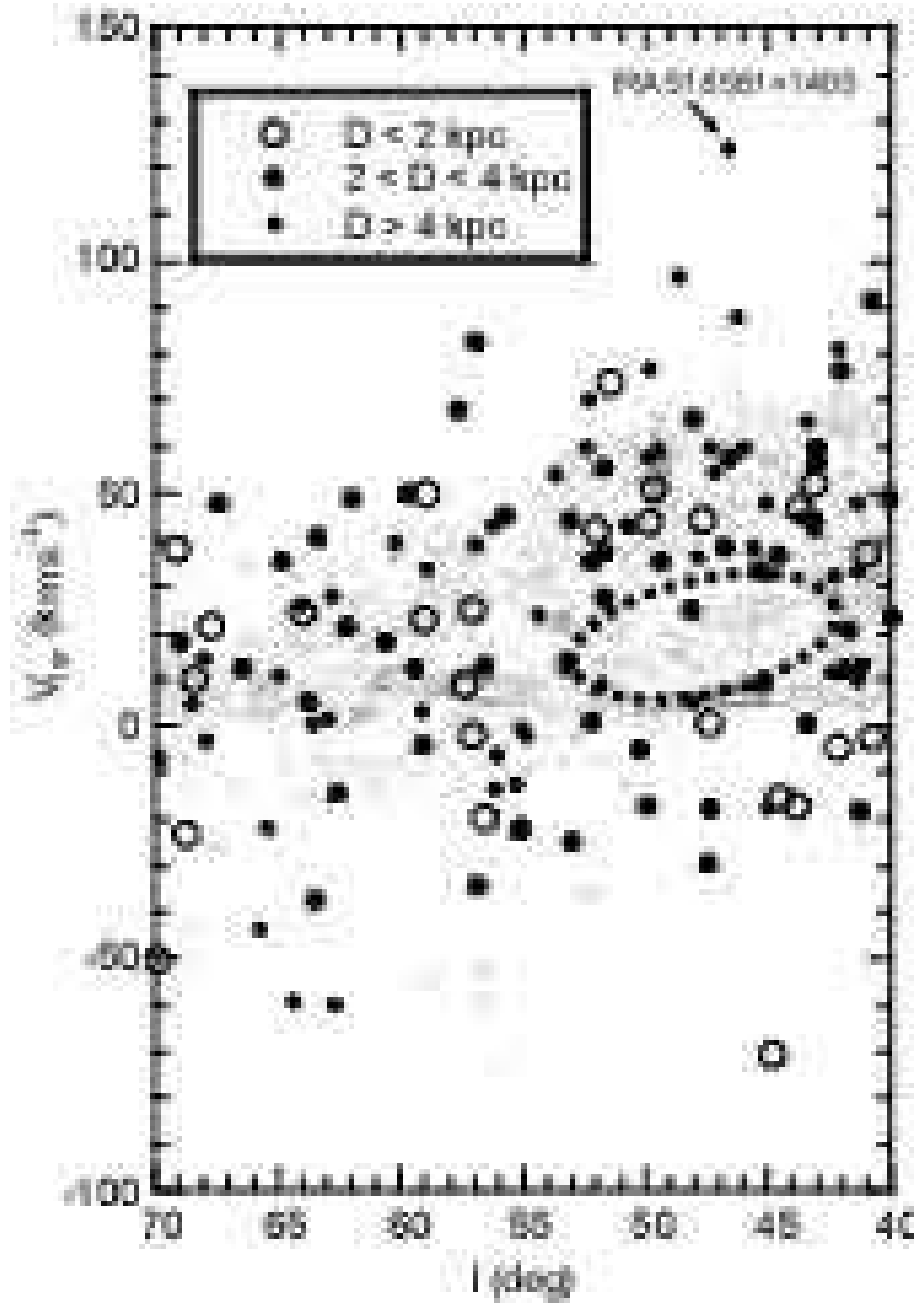

  \begin{center}
    \FigureFile(120mm,10mm){fig8.eps}
  \end{center}
  \caption{Longitude--velocity diagram of the SiO detected sources overlaid on the CO $v$--$l$ map, integrated in the range between $b=\pm3^{\circ}\hspace{-4.5pt}.\hspace{.5pt}25$ (taken from \cite{dam87}). The region surrounded by the dotted ellipse indicates a source vacant area (see text).
  }\label{fig:sample}
\end{figure}

\newpage
\renewcommand{\thefigure}{9}
\begin{figure}
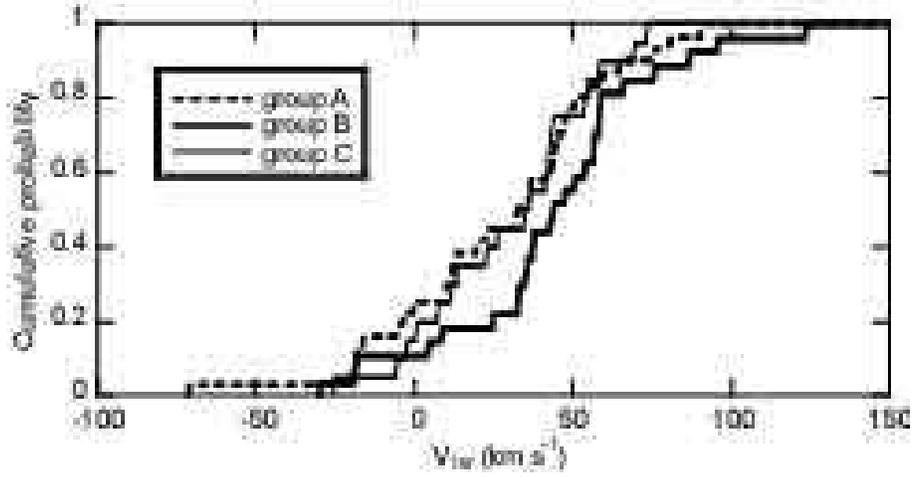

  \begin{center}
    \FigureFile(120mm,10mm){fig9.eps}
  \end{center}
  \caption{Cumulative probability plot of the $V_{\textrm{lsr}}$ distribution of the present sample. The cumulative probability is shown for three groups (see text).}\label{fig:sample}
\end{figure}

\newpage
\renewcommand{\thefigure}{10}
\begin{figure}
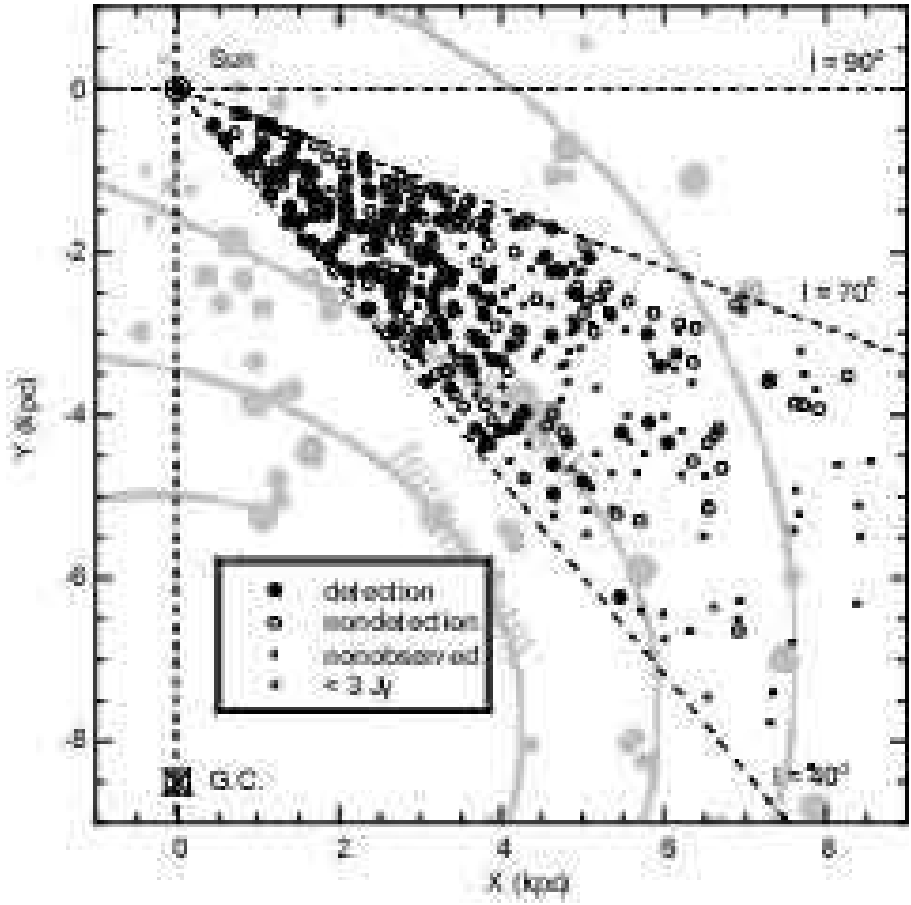

  \begin{center}
    \FigureFile(120mm,10mm){fig10.eps}
  \end{center}
  \caption{Positions of the observed sources projected onto the galactic plane, overlaid on the spiral model of the Galaxy \citep{tay93}. The positions of the Sun and the Galactic center are at the origin and $(X, Y)=(0, -8.5)$, respectively. }\label{fig:sample}
\end{figure}

\newpage
\renewcommand{\thefigure}{11}
\begin{figure}
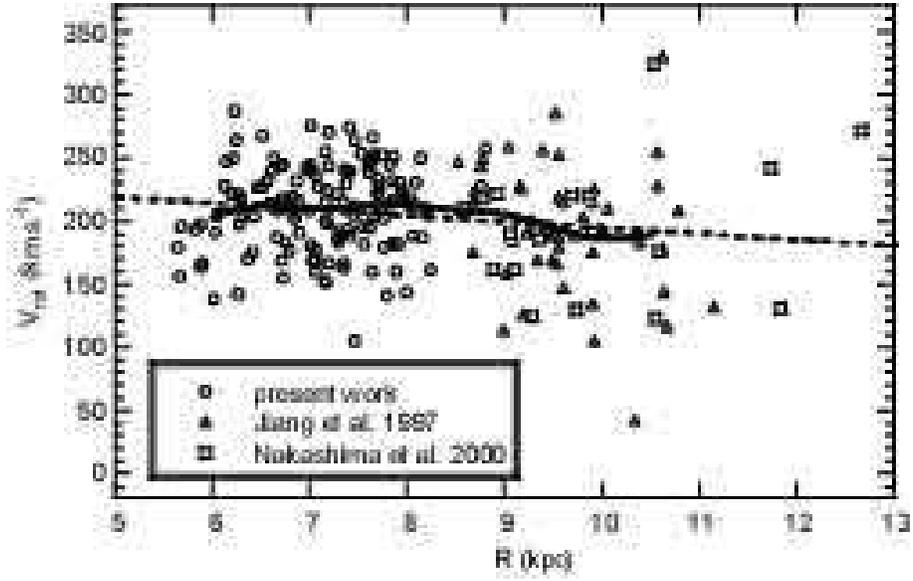

  \begin{center}
    \FigureFile(120mm,10mm){fig11.eps}
  \end{center}
  \caption{Rotation curve of SiO maser sources. The solid line indicates the running mean of the data.}\label{fig:sample}
\end{figure}



{\setlength{\tabcolsep}{3pt}\footnotesize
\begin{longtable}{lrrcrrrcrr}
\caption{List of detections.}
\hline\hline
 & & $J=1$--$0$ & $v=1$ &  & & $J=1$--$0$ & $v=2$ & &   \\
IRAS name & $V_{\textrm{lsr}}$ & $T_{\textrm{a}}$ & $S$ & r.m.s. & $V_{\textrm{lsr}}$ & $T_{\textrm{a}}$ & $S$ & r.m.s. & Obs. date \\
&\multicolumn{4}{c}{------------------------------------------------} & \multicolumn{4}{c}{------------------------------------------------}&\\ 
  & (km s$^{-1}$) & (K) & (K km s$^{-1}$) & (K) & (km s$^{-1}$) & (K) & (K km s$^{-1}$) & (K) & (yymmdd.d) \\
\hline
\endhead
\hline
\endfoot
\hline
\multicolumn{10}{l}{Note. The numbers in parentheses are tentative detections.}\\
\endlastfoot
18413+1354 & $-$12.9 & 13.364 & 47.770 & 0.084 & $-$17.9 & 5.335 & 31.542 & 0.099 & 000529.2 \\
18423+1616 & 35.9 & 0.595 & 2.216 & 0.052 & 40.8 & 0.495 & 1.368 & 0.071 & 000417.4 \\
18432+1343 & 37.1 & 0.712 & 2.323 & 0.058 & 37.2 & 0.878 & 2.116 & 0.080 & 000415.3 \\
18456+1351 & $-$17.3 & 0.730 & 1.428 & 0.054 & $-$17.7 & 0.516 & 0.895 & 0.064 & 000523.0 \\
18498+1014 & 81.5 & 0.149 & 0.395 & 0.029 & 80.8 & 0.185 & 0.524 & 0.034 & 000524.1 \\
18506+0912 & 47.8 & 0.225 & 0.250 & 0.043 & 48.1 & 0.334 & 0.541 & 0.052 & 000418.3 \\
18511+1141 & 45.0 & 0.101 & 0.361 & 0.027 & 44.8 & 0.180 & 0.028 & 0.033 & 000524.2 \\
18528+1543 & $-$18.3 & 1.293 & 6.896 & 0.122 & $-$16.6 & 0.835 & 4.528 & 0.157 & 000415.4 \\
\end{longtable}}


{\setlength{\tabcolsep}{3pt}\footnotesize
\begin{longtable}{lccr}
\caption{List of non-detections}
\hline\hline
IRAS name & r.m.s. ($J=$1--0, $v=$1) & r.m.s. ($J=$1--0, $v=$2) & Obs. date \\
 & ------------------------ & ------------------------ & \\ 
 & (K) & (K) & (yymmdd.d) \\
\hline
\endhead
\hline
\endfoot
\hline
\endlastfoot
18392+1328 & 0.074 & 0.052 & 000415.3 \\
18425+1727 & 0.057 & 0.041 & 000415.3 \\
18462+1208 & 0.069 & 0.045 & 000417.3 \\
18490+1158 & 0.044 & 0.030 & 000417.3 \\
18501+1019 & 0.091 & 0.069 & 000415.4 \\
18512+2029 & 0.091 & 0.069 & 000415.4 \\
18520+1014 & 0.067 & 0.050 & 000417.3 \\
18526+0945 & 0.060 & 0.053 & 000418.3 \\
\end{longtable}}


{\setlength{\tabcolsep}{3pt}\footnotesize
\begin{longtable}{lcrrrrr}
\caption{MSX counterparts of IRAS sources with large separations.}
\hline\hline
IRAS name & F$_{12}^{\textrm{iras}}$ & F$_{25}^{\textrm{iras}}$ & MSX name & F$_{12}^{\textrm{msx}}$ & F$_{21}^{\textrm{msx}}$ & Separation. \\
 & Jy & Jy & & Jy & Jy & (arcsec) \\
\hline
\endhead
\hline
\endfoot
\hline
\endlastfoot
19102+1356 & 5.78 & 5.71 & MSX5C\_G047.7990+01.7814 & 7.501  & 4.900  & 37 \\
19140+1313 & 14.27 & 13.25 & MSX5C\_G047.6006+00.6186 & 27.033  & 21.611  & 38 \\
19276+0736 & 6.36 & 4.03 & MSX5C\_G044.2837$-$04.9539 & 7.250  & 5.599  & 29 \\
19348+2136 & 50.25 & 33.37 & MSX5C\_G057.3350+00.3155 & 49.489  & 37.488  & 25 \\
19509+2930 & 13.92 & 13.84 & MSX5C\_G065.9956+01.1562 & 26.909  & 29.816  & 34 \\
\end{longtable}}


{\setlength{\tabcolsep}{3pt}\footnotesize
\begin{longtable}{lrrcrrrcrr}
\caption{Results of Confirmation at MSX positions}
\hline\hline
 & & $J=1$--0 & v=1 &  & & $J=1$--0 & v=2 & &   \\
IRAS name & $V_{\textrm{lsr}}$ & $T_{\textrm{a}}$ & $S$ & r.m.s. & $V_{\textrm{lsr}}$ & $T_{\textrm{a}}$ & $S$ & r.m.s. & Obs. date \\
&\multicolumn{4}{c}{------------------------------------------------} & \multicolumn{4}{c}{------------------------------------------------}&\\ 
  & (km s$^{-1}$) & (K) & (K km s$^{-1}$) & (K) & (km s$^{-1}$) & (K) & (K km s$^{-1}$) & (K) & (yymmdd.d) \\
\hline
\endhead
\hline
\endfoot
\hline
\endlastfoot
19102+1356 & 52.1 & 0.580 & 1.403 & 0.063 & 52.6 & 0.521 & 0.994 & 0.082 & 010319.3 \\
19140+0652 & 27.3 & 0.366 & 0.886 & 0.041 & 28.3 & 0.287 & 0.844 & 0.052 & 010319.3 \\
19276+0739 & 94.9 & 0.499 & 3.040 & 0.055 & 93.2 & 0.801 & 2.391 & 0.078 & 010319.3 \\
19348+2136 & 34.3 & 0.134 & 0.831 & 0.026 & 34.8 & 0.123 & 0.265 & 0.035 & 010319.4 \\
19509+2930 & $-$40.0 & 2.194 & 5.043 & 0.132 & $-$39.5 & 2.765 & 5.141 & 0.165 & 010523.0 \\
\end{longtable}}


{\setlength{\tabcolsep}{2pt}\scriptsize
\begin{longtable}{lrrrrrrrlll}
\caption{IRAS Data.}
\hline\hline
IRAS name & $l$ & $b$ & $F_{12}$ & $C_{12}$ & $C_{23}$ & $D_{L}$ & $V_{\textrm{lsr}}$ & OH & H$_2$O & SiO \\
 & ($^\circ$) & ($^\circ$) & (Jy) &  & & (kpc) & (km s$^{-1}$) \\
\hline
\endhead
\hline
\endfoot
\hline
\multicolumn{11}{l}{References. --- 
1. \citet{ale89}; 
2. \citet{ben96}; 
3. \citet{ces88}; }\\
\multicolumn{11}{l}{
4. \citet{che93}; 
5. \citet{com90}; 
6. \citet{ede88}; }\\
\multicolumn{11}{l}{
7. \citet{eng89}; 
8. \citet{eng96}; 
9. \citet{eng86}; 
}\\
\multicolumn{11}{l}{
10. \citet{eng88}; 
11. \citet{gal89}; 
12. \citet{gom90};  
}\\
\multicolumn{11}{l}{
13. \citet{hal90}; 
14. \citet{jew91}; 
15. \citet{jew85}; 
}\\
\multicolumn{11}{l}{
16. \citet{jia99}; 
17. \citet{les92}; 
18. \citet{lew94}; 
}\\
\multicolumn{11}{l}{
19. \citet{lew93}; 
20. \citet{lew95}; 
21. \citet{lew90}; 
}\\
\multicolumn{11}{l}{
22. \citet{mar99}; 
23. \citet{nym98}; 
24. \citet{sea95}; 
}\\
\multicolumn{11}{l}{
25. \citet{siv90}; 
26. \citet{slo85}; 
27. \citet{tel91}; 
}\\
\multicolumn{11}{l}{
28. \citet{tel89};
29. \citet{lew97};
30. \citet{lew87}
}\\
\multicolumn{11}{l}{
31. \citet{tak00}
}\\
\endlastfoot
18392+1328 & 43.95 & 8.28 & 8.968 & $-$0.08 & $-$0.82 & 4.1 & $\cdots$ & & & \\
18413+1354 & 44.56 & 8.02 & 225.100 & $-$0.17 & $-$0.85 & 0.6 & $-$15.4 & \phantom{00}y$^{26}$ & \phantom{00}y$^{5}$ & \phantom{00}y$^{7}$ \\
18423+1616 & 46.82 & 8.85 & 8.255 & $-$0.17 & $-$0.82 & 3.3 & 38.4 & \phantom{00}n$^{21}$ & & \\
18425+1727 & 47.91 & 9.32 & 24.970 & $-$0.19 & $-$0.72 & 1.8 & $\cdots$ & \phantom{00}n$^{21}$ & & \\
18432+1343 & 44.60 & 7.51 & 15.230 & $-$0.13 & $-$0.80 & 2.7 & 37.1 & \phantom{00}n$^{21}$ & & \\
18456+1351 & 44.99 & 7.05 & 3.056 & $-$0.14 & $-$0.53 & 5.9 & $-$17.5 & \phantom{00}n$^{21}$ & & \\
18462+1208 & 43.52 & 6.15 & 8.551 & $-$0.18 & $-$0.98 & 3.1 & $\cdots$ & \phantom{00}n$^{4}$ & & \\
18490+1158 & 43.67 & 5.46 & 9.061 & $-$0.02 & $-$0.73 & 4.8 & $\cdots$ & \phantom{00}y$^{17}$ & \phantom{00}n$^{8}$ & \\
\end{longtable}}





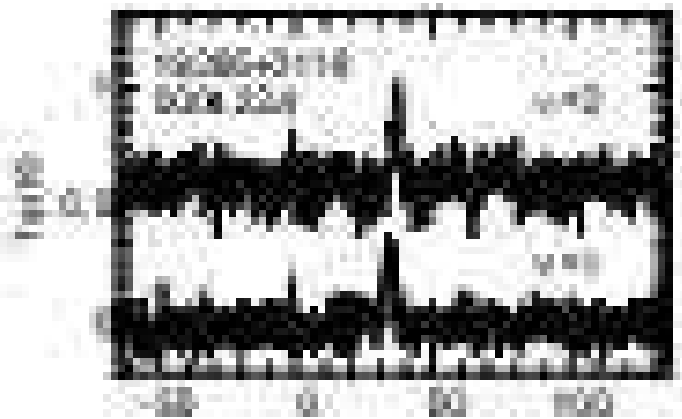
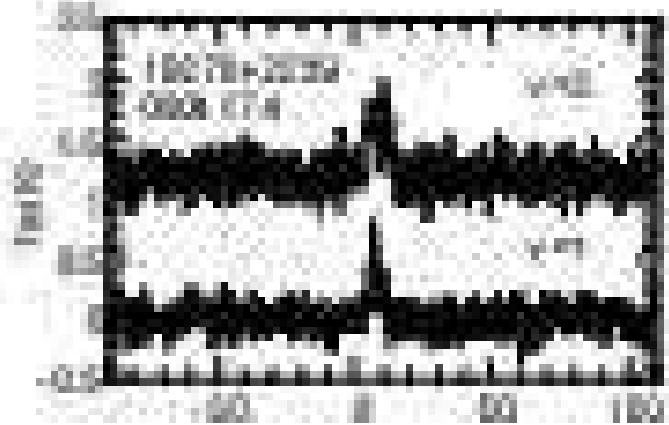
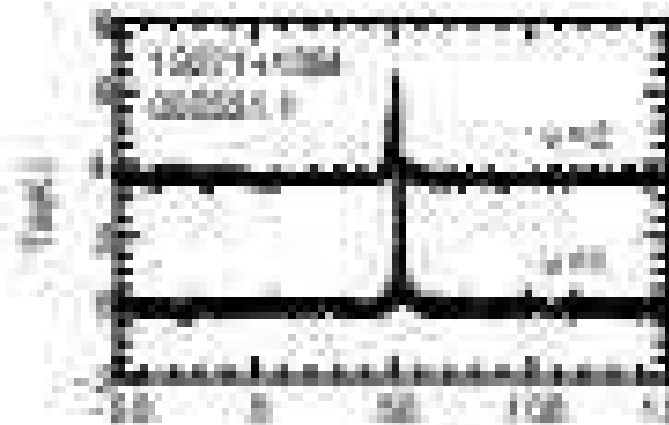
\end{document}